\documentclass[sigconf,authorversion=true]{acmart}

\usepackage{tabularx}

\usepackage{booktabs}
\usepackage[nolist]{acronym}

\usepackage{xspace}

\newcommand{\trec}{\ac{TREC}\xspace}
\newcommand{\trecpm}{\ac{TREC-PM}\xspace}
\newcommand{\infndcg}{\ac{infNDCG}\xspace}
\newcommand{\pubmed}{\textsc{PubMed}\xspace}
\newcommand{\ctdotorg}{\textsc{ClinicalTrials.org}\xspace}

\begin{acronym}[SNOMED CT]  %
    \acro{CUI}{Concept Unique Identifier}
    \acro{ES}{\textsc{ElasticSearch}}
    \acro{BA}{Biomedical Abstracts}
    \acro{CT}{Clinical Trials}
    \acro{FOSS}{Free and open-source software}
    \acro{GS}{Gold Standard}
    \acro{IR}{Information Retrieval}
    \acro{infNDCG}{inferred normalized discounted cumulative gain}
    \acro{PM}{Precision Medicine}
    \acro{LTR}{Learning-to-Rank}
    \acro{ML}{Machine Learning}
    \acro{MAP}{Mean Average Precision}
    \acro{MeSH}{Medical Subject Headings}
    \acro{NCBI}{National Center for Biotechnology Information}
    \acro{NDCG}{normalized discounted cumulative gain}
    \acro{NLP}{Natural Language Processing}
    \acro{NIST}{National Institute of Standards and Technology}
    \acro{TREC}{Text REtrieval Conference}
    \acro{TREC-PM}{TREC Precision Medicine}
    \acro{TF-IDF}{Term Frequency - Inverse Document Frequency}
    \acro{UIMA}{Unstructured Information Management Architecture}
    \acro{UMLS}{Unified Medical Language System}
    \acro{PFS}{progression free survival}
    \acro{OS}{overall survival}
    \acro{SMBO}{sequential model-based optimization}
    \acro{SMAC}{Sequential Model-based Algorithm Configuration}
\end{acronym}

\AtBeginDocument{%
  \providecommand\BibTeX{{%
    \normalfont B\kern-0.5em{\scshape i\kern-0.25em b}\kern-0.8em\TeX}}}

\copyrightyear{2020} 
\acmYear{2020} 
\setcopyright{acmlicensed}
\acmConference[SIGIR '20]{Proceedings of the 43rd International ACM SIGIR Conference on Research and Development in Information Retrieval}{July 25--30, 2020}{Virtual Event, Xi'an, China}
\acmBooktitle{Proceedings of the 43rd International ACM SIGIR Conference on Research and Development in Information Retrieval (SIG\-IR '20), July 25--30, 2020, Virtual Event, Xi'an, China}
\acmPrice{15.00}
\acmDOI{10.1145/3397271.3401048}
\acmISBN{978-1-4503-8016-4/20/07}

\begin{document}

\fancyhead{}

\title{What Makes a Top-Performing Precision Medicine Search Engine? Tracing Main System Features in a Systematic Way}

\author{Erik Faessler}
\orcid{0003-1193-5103}
\affiliation{%
  \institution{Jena University Language and
Information Engineering (JULIE) Lab\\
Friedrich-Schiller-Universit{\"a}t Jena}
  \city{Jena}
  \country{Germany}
}
\email{erik.faessler@uni-jena.de}

\author{Michel Oleynik}
\affiliation{%
  \institution{Institute for Medical Informatics, Statistics and Documentation,\\
  Medical University of Graz}
  \city{Graz}
  \country{Austria}}
\email{michel.oleynik@stud.medunigraz.at}
\orcid{0001-9099-6259}

\author{Udo Hahn}
\affiliation{%
  \institution{Jena University Language and
Information Engineering (JULIE) Lab\\
Friedrich-Schiller-Universit{\"a}t  Jena}
  \city{Jena}
  \country{Germany}
}
\email{udo.hahn@uni-jena.de}

\begin{abstract}
From 2017 to 2019 the Text REtrieval Conference (TREC) held a challenge task on precision medicine using documents from medical publications (PubMed) and clinical trials.
Despite lots of performance measurements carried out in
these evaluation campaigns, the scientific community is still pretty unsure about the impact individual system features and their weights have on the overall system performance.
In order to overcome this explanatory gap, we first determined optimal feature configurations using the Sequential Model-based Algorithm Configuration (SMAC) program and applied its output to a BM25-based search engine.
We then ran an ablation study to systematically assess the individual contributions of relevant system features:
BM25 parameters, query type and weighting schema, query expansion, stop word filtering, and
keyword boosting.
For evaluation, we employed the gold standard data from the three \trecpm\ installments to evaluate the effectiveness of different features using the commonly shared  infNDCG metric.
\end{abstract}

\begin{CCSXML}
<ccs2012>
<concept>
<concept_id>10002951.10003317</concept_id>
<concept_desc>Information systems~Information retrieval</concept_desc>
<concept_significance>500</concept_significance>
</concept>
<concept>
<concept_id>10002951.10003317.10003318.10003321</concept_id>
<concept_desc>Information systems~Content analysis and feature selection</concept_desc>
<concept_significance>300</concept_significance>
</concept>
<concept>
<concept_id>10002951.10003317.10003359.10003362</concept_id>
<concept_desc>Information systems~Retrieval effectiveness</concept_desc>
<concept_significance>100</concept_significance>
</concept>
<concept>
<concept_id>10002951.10003317.10003371</concept_id>
<concept_desc>Information systems~Specialized information retrieval</concept_desc>
<concept_significance>100</concept_significance>
</concept>
<concept>
<concept_id>10010405.10010444.10010449</concept_id>
<concept_desc>Applied computing~Health informatics</concept_desc>
<concept_significance>100</concept_significance>
</concept>
</ccs2012>
\end{CCSXML}

\ccsdesc[500]{Information systems~Information retrieval}
\ccsdesc[300]{Information systems~Content analysis and feature selection}
\ccsdesc[100]{Information systems~Retrieval effectiveness}
\ccsdesc[100]{Information systems~Specialized information retrieval}
\ccsdesc[100]{Applied computing~Health informatics}

\keywords{TREC, precision medicine, search engine evaluation}

\maketitle

\section{Introduction}

According to the U.S.\ National Research Council \cite{NAP13284}, the goal of \textit{``precision medicine''} is to tailor patients' treatment according to their individual characteristics, thus providing them with the best available care. While physicians were
traditionally trained
how to treat a disease following \textit{general} principles and rules, today they need to know how to treat a disease in a \textit{specific} human subject. This causes an immense increase of the number of variables doctors have to account for when creating treatment plans, calling for measures to effectively and efficiently provide the required information.

To make such information available, text retrieval engines are an obvious choice.
Yet, biomedical \ac{IR} faces unique challenges owing to an unclear translation of information needs into queries and a non-consensual notion of relevance.
For instance, while it is acceptable to retrieve documents about any of colon, rectum, or anal cancer for the query \textit{``colorectal cancer''}, a query about \textit{``cholangiocarcinoma''} should strictly restrict documents to ones about the biliary duct between the gallbladder and the duodenum.
The biomedical domain also exhibits specific terminology (such as \textit{``HER2''}, the human epidermal growth factor receptor 2), showing a large degree of lexical variation and ambiguity (e.g., \textit{``HER2''} is synonym to \textit{``ERBB2''}, the erythroblastic oncogene B 2).

For proper system engineering, the \ac{IR} community needs document collections for system evaluation, but such datasets, including query-specific relevance judgments, are still quite rare for text-based \ac{PM}. The earliest work is due to \citet{Hersh94} who developed \textsc{OHSUMed}. This corpus consists of a subset of clinical \textsc{Medline} abstracts spanning the years 1987--1991, 106 topics from clinicians, and an accompanying set of relevance judgments. Patient-related data is also available from the \textsc{ShARe/CLEF eHealth} Evaluation Lab \cite{Kelly14}, yet its focus is more on patients' understandability of clinical records and information from the Web. The most recent and commonly shared dataset for \ac{PM}-related retrieval studies, however, is provided by the organizers of the \ac{TREC}.\footnote{https://trec.nist.gov/overview.html}

\ac{TREC} has featured life science-focused tracks for quite a while, ranging initially from bioinformatics issues to clinical topics more recently. 
Starting in 2014, the \ac{TREC} Clinical Decision Support Track (CDS) \cite{Simpson14,Roberts15,Roberts16} focused on the retrieval of relevant biomedical articles to answer clinical questions related to 
medical records. This track was superseded in 2017 by the \ac{TREC-PM} track \cite{Roberts17,Roberts18,Roberts19}, with its focus on personalized patient treatment within the framework of the emerging ``precision medicine'' paradigm \cite{collins2015new}.
\ac{TREC-PM} focused on two tasks, namely the retrieval of relevant (1)  biomedical articles from \textsc{PubMed} and (2) clinical trials from the \texttt{ClinicalTrials.gov} collection composed of
synthetic oncology patient cases which contain mainly information about the disease, biomarkers, and demographics of a person.

For all three \ac{TREC-PM} editions, submissions were limited to five different sets of results (runs) per task and participating team. The respective gold standards were created after the submission phase 
by result list pooling. Thus, the final scores for all submissions are known. 
But despite the organizers' efforts to isolate major success factors in the top-performing systems in their survey papers \cite{Roberts17,Roberts18,Roberts19},
up until now no consensus could be found that would explain exactly which features made the highest-ranked systems perform so well.

The fragmentary and inconclusive nature of our knowledge about how to build effective \ac{IR} systems indicates a
lack of solid systems engineering foundations reflecting the validated state of the art in this field.
To fully explore the features that a good PM search engine should have, the systematic assessment of a large set of hyperparameters is necessary. This is a very challenging problem for computer science, in general, and \ac{IR}, in particular.

Virtually every information system dealing with unstructured data comes with free-to-choose parameters (features of choice, their weights in the overall decision process, and cut-offs as brute-force decision criteria) whose contribution to the final outcome is rarely investigated in depth---they come as ``experience''-based defaults or are varied in a mostly shallow way to motivate specific  choices (e.g., cut-offs based on thin experimental evidence of the system designers gathered in the parameter-setting phase of system development). Hence, it is often hard to decide whether the underlying algorithm or intuitively (well-)tuned parameters that go into it are the source of success in evaluation experiments.
As an anecdotal evidence for these claims, \citet{Cox11} explored a single, richly parameterized model family, yielding classification performance that ranged from chance to state-of-the-art performance depending solely on hyperparameter choices.

This explanatory gap motivates this paper.
We propose a methodologically sound approach to evaluate a set of system features by (1) tuning the free (hyper-)parameters that every search engine has in an effort to find good, if not optimal, system configurations;  (2) running an ablation study of the best configurations to gain deeper insights into the contributions of individual system features.

We apply this approach to an existing \ac{PM} search engine extended with promising features taken from the \ac{TREC-PM} literature.
Hence, rather than configuring and testing a new state-of-the-art system, we here strive for finding the most influential system features that determine the current state of the art. We stick to rather simple features in an effort to find a stable set of core features and parameters that should be explanatory of how to perform well on the existing datasets and may serve as an experimentally grounded (and systematically validated) reference feature set for future work.

\section{Related Work}\label{sec:rel}

Previous research already identified the need to consistently assess \trec results so that conflicting claims could be solved and experimental evidence be generalized to further advance the field.
Building on experience from the \ac{TREC}-CDS track,
\citet{Nguyen18} 
built a platform that allows to compare different document and query processing techniques using
different search parameters. 
Continuing on this line, \citet{Nguyen19} offer a platform for common experimentation with \trecpm data and used it to benchmark several variants and combinations proposed by leading teams. Their system supports terminology-based query expansion (related to genes and diseases), ranking models (different variants of \ac{LTR} and BM25) and re-ranking strategies based on citation analysis.
Similarly, \citet{Chen18} also proposed a common framework to evaluate methods (keyphrase extraction, query expansion, and supervised results re-ranking) developed for TREC-PM, but they neither made their system publicly available, nor provided comparison results.

The issue of query expansion by terminological resources has also been explored by \citet{Stokes08}; using data from the TREC 2006 Genomics Track task \cite{Hersh06}, they improved the \textsc{Okapi} baseline passage \ac{MAP} performance by 185\% with parameter juggling.
In particular, they show, first, that the main single factor affecting the accuracy of the retrieval process is the ranking metric being used (comparing with the standard  \textsc{Okapi} ranking algorithm) and, second, that the expansion with synonyms and
lexical variants is much more effective than the inclusion of hierarchical terms (taxonomies) from ontologies.

Studying parametric retrieval functions, i.e.,
investigating near-optimal choices of their free parameters, has become a relevant topic of research both in \ac{ML} and \ac{IR} in the past years. \citet{Bergstra13} presented a meta-modeling approach that replaced hand-tuning of configuration parameters (hyperparameters) with a reproducible and unbiased Bayesian optimization process.
The authors implemented a broad class of image feature extraction and classification models to formalize the steps of selecting the model parameters and evaluated it on three disparate computer vision problems. They compared random search in that model class with a more sophisticated algorithm for hyperparameter optimization (Tree of Parzen Estimators \cite{Bergstra2011}) and found that the optimization-based search strategy not only outperformed random search but also recovered or improved on the best known (hand-tuned) configurations for all three image classification tasks.

In order to systematically explore the huge search spaces of (optimal) hyperparameter settings, robust search machinery is needed. A comparison of three Bayesian approaches to hyperparameter optimizers, \textsc{Hyperopt}, \textsc{Spearmint} \cite{Snoek12} and \textsc{SMAC} \cite{Hutter11}, was provided by \citet{Eggensperger2013}, with experimental evidence indicating the superiority of \textsc{SMAC} over its two alternatives. For our configuration search experiments (cf.\ Section \ref{sec:smac}), we therefore selected  \ac{SMAC} as parameter tuning engine.
Still, there is further progress in the development of efficient tools (cf., e.g., \cite{Golovin17,Sievert19}) and methodological approaches going beyond Bayesian optimization \cite{Falkner18,Li18}.

Perhaps the earliest account of hyperparameter tuning dedicated to parametric retrieval information functions is due to \citet{He03} who described a specific query-focused approach to term frequency normalization parameter tuning, i.e., focus is on one single parameter only. Subsequently, \citet{Taylor06} applied greedy line searches (described in detail by \citet{Costa18a}) and an extension of the gradient descent approach to training sets of up to 2,048 queries, testing the impact of up to 375 parameters on \ac{NDCG} scores \cite{Jarvelin02}.
Only recently, \citet{Ghawi19} performed hyperparameter tuning (for three hyperparameters) using grid search for a text categorization task employing a kNN algorithm with BM25 similarity \cite{Robertson10}.
\citet{Ghawi19} also considered the role of document length in their experiments and found that tuning methods yield higher parameter values on longer documents compared with shorter ones.
An additional observation they made is that their method became faster with larger grids and longer documents (see Section \ref{sec:disc} for our observations on the effect of document length).
Since grid search exhaustively enumerates all combinations of hyperparameters and evaluates each combination, this approach is limited to small-sized hyperparameter sets and does not scale well.

\citet{Costa18a} treated the parameter tuning problem as a mathematical optimization of retrieval functions with a  black-box 
optimization approach based on surrogate models (i.e., assuming a computable, but not analytically available objective function, e.g., the optimization of variants of BM25). Technically, they employed a variant of the Metric Stochastic Response Surface Method \cite{Regis07}
implemented in the open-source library \textsc{RBFOpt} \cite{Costa18b}.
Their approach yields near-optimal results for the 
(2T + 1)-parameter version of BM25F (T being the number of stream weights) \cite{Robertson04} and the two hyperparameters of BM25 \cite{Robertson10} (experiments are reported for up to nine free parameters for BM25F); it also outperformed classical line and grid search (Bayesian optimization was not considered). 

Our work complements this previous work with a broader set of parameter types (including BM25, query types, weighting schemata, stop word filtering, and keyword boosting) that also pay tribute to the specific \ac{PM} problem domain (via disease and gene expansion). Using the hyperparameter optimization as a solid technical framework for \textit{optimizing} single parameters from these types, we then try to \textit{explain} their single contribution to the performance figures of a \ac{PM} search engine via an elaborate ablation study.

\section{Methods}

\subsection{Data}
\label{sec:data}

\begin{figure}[h]
\centering
    \includegraphics[width=0.4\textwidth]{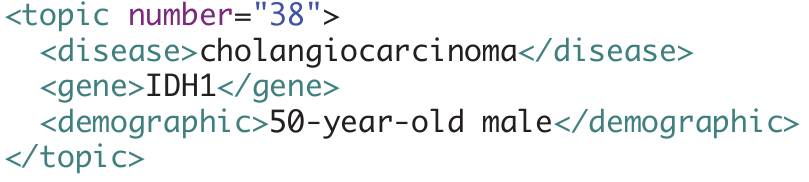}
\caption{Example of a \trecpm topic.}
\Description{A snippet of a XML file. It shows that the topic number 38 has disease cholangiocarcinoma, gene IDH1, and demographic 50-year-old male.}
\label{img:trec-topic}
\end{figure}

We employed the datasets from the \trecpm tracks held between 2017 and 2019 (see Table~\ref{tab:trec_data}). The document corpora used for the \ac{BA} and \ac{CT} tracks were snapshots of  \pubmed and \ctdotorg, respectively. For the years 2017 and 2018 the document collections were the same; in 2019, more recent versions of the collections were provided. 
Each year, a new set of queries, called topics (see Figure~\ref{img:trec-topic} for an example), were released to the participants.
In total, 120 topics were formulated for both the \ac{BA} and \ac{CT} tasks and manually labeled subsets were created each year summing up to 63,387 and 40,625 relevance assessments, respectively.

\begin{table}%
\caption{TREC-PM data overview.}
\label{tab:trec_data}
\begin{tabular}{lll}
    \toprule
    & Biomedical Abstracts & Clinical Trials \\
	\midrule
	Topics & 120 & 120 \\
	Documents & 29,137,141 & 306,238 \\
	Relevance assessments & 63,387 & 40,625 \\
 	\quad Definitely relevant & 8,035 (12.70\%) & 1,794 (4.40\%) \\
	\quad Partially relevant & 6,972 (11.00\%) & 3,609 (8.90\%) \\
	\bottomrule
\end{tabular}
\end{table}

To perform the experiments reported in this paper, we extended an evaluation framework created by participating teams \citep{lopezgarcia2017trec,oleynik2018hpi,faessler2019julie} in the course of the three TREC-PM editions. 
It employs \ac{ES}\footnote{In version 5.4, available at \url{https://www.elastic.co}.} as the index and search server. For all our experiments, we used the default functionality of \ac{ES} without any custom extensions.
We enriched documents with gene mention annotations produced by the \textsc{Banner} gene tagger\footnote{\url{http://banner.sourceforge.net}} trained on data from the BioCreative II Gene Mention task.\footnote{\url{http://biocreative.sourceforge.net/biocreative\_2\_gm.html}}
Our source code is available at \url{https://doi.org/10.5281/zenodo.3856403} under the MIT license.

\subsection{Query Layout}
\label{sec:querylayout}

\begin{figure}[h]
\centering
    \includegraphics[width=0.45 \textwidth]{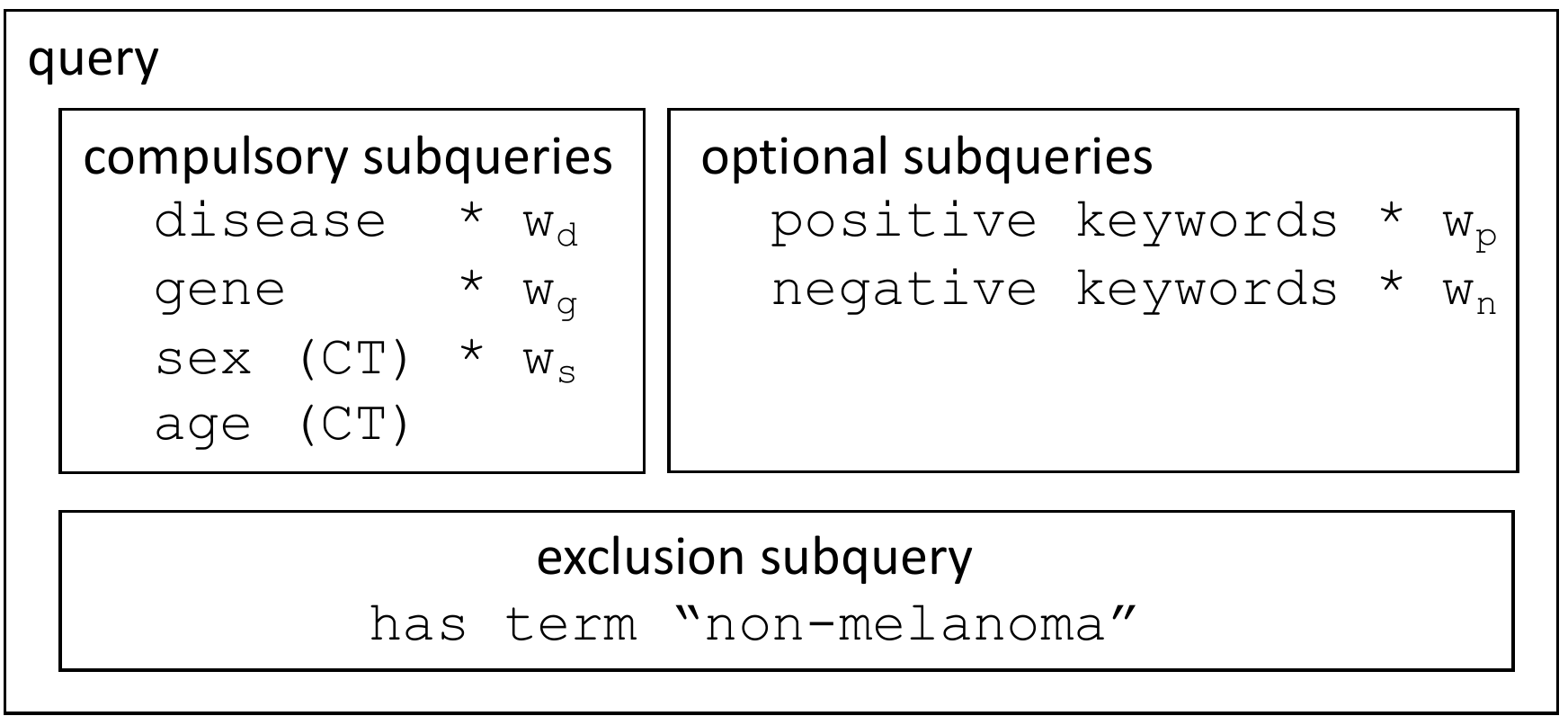}
\caption{A schema of the query structure used throughout this work. Subqueries marked with \ac{CT} are only used when searching clinical trials. 
Weights $w_x$  are assigned to the subqueries where each subquery has its own weight.}
\Description{Figure 2. Fully described in the text.}
\label{img:query}
\end{figure}

As the \trecpm topics needed to be translated to \ac{ES} queries for result list retrieval, we tried to stick to two simple questions: (1) \emph{What is required
to find relevant documents?} and (2) \emph{How can found, yet irrelevant, documents be pushed towards the end of the result list?} Figure \ref{img:query} illustrates the main shape of the queries we created for document retrieval.

The \ac{TREC-PM} topic aspects are assembled in a compulsory compound query to restrict the results to potentially relevant results and thus address the first question. We used 
demographic information only for clinical trials, since the structured data explicitly contain
such relevant pieces of information, whereas \pubmed abstracts cannot easily be  matched to the demographic aspect.

For the second question, a set of optional subqueries provided additional relevance signals aiming to match general aspects of precision medicine to distinguish between \ac{PM} and non-\acs{PM} documents matched by the compulsory query part. We leveraged positive and negative PM keywords to tackle this issue at the lexical level.

\begin{figure}[h]
\centering
    \includegraphics[width=0.4\textwidth]{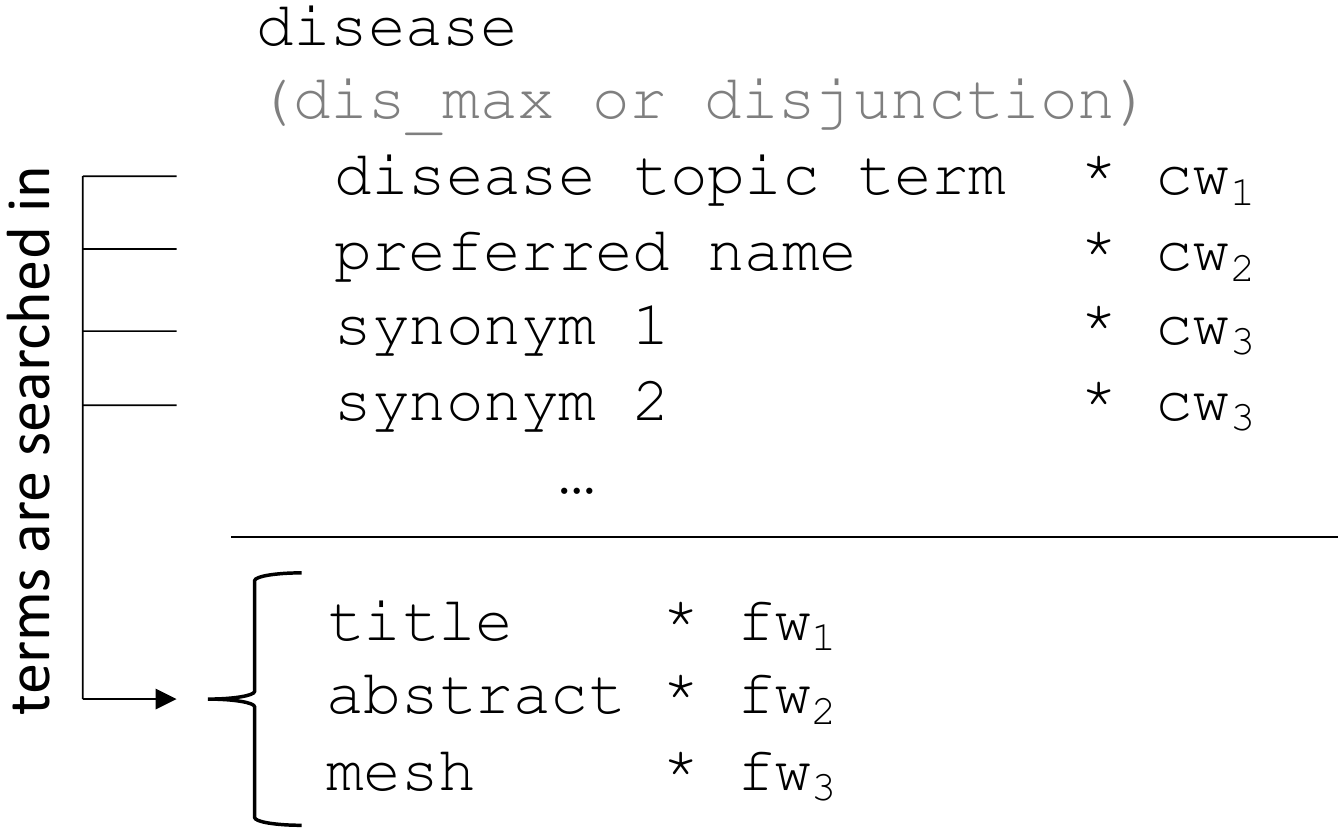}
\caption{The structure of the \textit{disease} subquery. The terms associated with the value of the disease topic aspect are searched in a compound query in the title, abstract, and MeSH fields. We explore simple Boolean disjunctions and the disjunction max compound query types. 
Weights $cw_i$ are assigned to the query clauses where each clause type has its own weight. Additionally, 
weights $fw_j$ are assigned to the fields in which terms are searched. The \textit{gene} subquery has a similar form but adds the \emph{gene} field with a weight of its own.}
\Description{Figure 3. Fully described in the caption.}
\label{img:diseasequery}
\end{figure}

Except for the \emph{age} subquery (a range filter) and the exclusion subquery, we assigned weights to all subqueries to be optimized by the hyperparameter search algorithm described in Section \ref{sec:smac}. Moreover, it can entirely disable subqueries should they turn out not to be effective at all.
The \emph{disease} and \emph{gene} subqueries are complex queries that encompass the terms from the topic description and the potential query expansion terms as illustrated in Figure~\ref{img:diseasequery} (see Section~\ref{sec:dismax} for a description of query types).

\subsection{Evaluation Strategy}
\label{sec:evalstrategy}

We report all evaluation results using the \infndcg score \cite{Yilmaz08}.\footnote{Note that samples of unjudged documents are not available for \ac{CT} in 2017.} Not only does \infndcg consider the exact relevance score --- 0 (not relevant), 1 (partially relevant), and 2 (definitively relevant) --- but it also accounts for the incompleteness of the gold standard data.
The scores were calculated using all available 120 \trecpm topics\footnote{90 for \ac{CT} due to the inability to calculate the infNDCG score for 2017.} by computing the arithmetic mean over all individual topic scores for a given system configuration.

To mitigate the risk of overfitting, we carried out 10-fold cross-validation across topics.
We uniformly balanced the distribution of diseases and genes across the partitions to avoid a bias resulting from topic clustering.\footnote{For instance, the topic disease aspect \emph{melanoma} appears 28 times in the data which would cause the parameter optimization to focus too much on this single disease in splits where it would appear disproportionately often.}
We thus carried out 20 independent parameter searches (following the remarks in Section~\ref{sec:smac}) to find good parameters for all training sets.

\subsection{Configuration Search}
\label{sec:smac}

Together with the BM25 ranking function in use (which contains two hyperparameters on its own) the query layout described in Section~\ref{sec:querylayout} exposes a large set of additional parameters, for most of which no \textit{ad hoc} best choice is known (see Table~\ref{tab:ranges}).
The complete search space is comprised of 100 parameters of which 55 were binary (i.e., they switch a feature on or off), 11 were categorical (e.g., for query types), and 34 were numerical.

\begin{table}%
  \caption{Configuration search: input feature space.}
  \label{tab:ranges}
  \begin{tabular}{lll}
    \toprule
    Configuration & Default & Input  \\
    \midrule
    BM25 Parameters\\
    \quad $b$ & 0.75 & $[0,1]$\\   %
    \quad $k_1$ & 1.20 & $[0,2]$ \\
    \midrule  %
    Query type \\
    \quad Expansions & disjunction & \{dis\_max, disjunction\} \\ %
    \quad Multi-word & bag-of-words & \{phrase, bag-of-words\} \\
    \midrule
    Weighting schema \\
    \quad Fields & 1.00 & $[0,3]$ \\
    \quad Clauses & 1.00 & $[0,3]$ \\
    \midrule
    Disease expansion \\
    \quad Preferred term & No & \{Yes, No\} \\    %
    \quad Synonyms & No & \{Yes, No\} \\    %
    \quad Hypernyms & No & \{Yes, No\} \\   %
    \quad Solid tumor & No & \{Yes, No\} \\
    \midrule
    Gene expansion \\
    \quad Synonyms & No & \{Yes, No\} \\ %
    \quad Description & No & \{Yes, No\} \\  %
    \quad Family & No & \{Yes, No\} \\
    \midrule
    Stop word filtering & No & \{Yes, No\} \\
    \midrule
    Keyword boosting \\
    \quad Positive & $\varnothing$ & \{words from Table~\ref{tab:pm-keywords}\} \\
    \quad Negative & $\varnothing$ & \{words from Table~\ref{tab:pm-keywords}\} \\
    \quad Non-melanoma & No & \{Yes, No\} \\
    \quad Gene tagger & 0.00 & $[0,3]$ \\ %
    \bottomrule
\end{tabular}
\end{table}

Since an exhaustive search across this huge parameter space
would have consumed prohibitive
time and computing resources,
we  employed \ac{SMAC},\footnote{\url{https://www.cs.ubc.ca/labs/beta/Projects/SMAC/}}
a technique that alternates between (a) obtaining scores from calls to some black-box algorithm and (b) fitting a random forest model to estimate the algorithm score for unseen parameter configurations.
All possible combinations were thus potentially considered by the algorithm.
For each cross-validation fold, the optimization algorithm ran independently from other splits and only on the respective training partitions. Thus, we obtained 20 optimized configurations.
Even though \ac{SMAC} can optimize parameters from scratch, we provided starting configurations to speed up the computation process.
Except for the underlined words from Table~\ref{tab:pm-keywords} (known to improve results based on previous experiments) and title weights (set to 2.00), we used default values from Table~\ref{tab:ranges}.

\subsection{Ablation Study}
\label{sec:ablation}

In order to assess which features contributed the most to the best \infndcg scores we achieved, we carried out an ablation study.
For each cross-validation split $i$, we used the best configuration we found on the training part $train_i$ and applied it to the test part $test_i$ as the baseline test score for this split.
Then, for each explored ablation feature group $ablation_j$, we calculated the \infndcg score for the baseline configuration modified by disabling the feature (group) (or setting it to the default value), which resulted in the score $s_{ij}$ of the $test_i$ partition for the $j$th ablation group. The final score was obtained by averaging across the tests splits. We report the score of $ablation_j$ as in Equation~(\ref{eq:ablation}), 
in which $N$ is the number of cross-validation splits. The reported baseline scores were obtained in a similar manner  by using the best configuration we found on $train_i$ applied to $test_i$ without manipulating the configurations.

\begin{equation}\label{eq:ablation}
score(ablation_j) := \frac{1}{N}\sum_{i=1}^{N} s_{ij}
\end{equation}

\subsection{Statistical Significance Testing}

For the statistical analysis of our results we used an approximate randomization test. The employed test statistic is the mean over the samples, making this a randomized version of the one-sample t-test \cite{cohen95}. This testing framework is suitable to find differences in the behavior of two systems. The basic idea is that when two systems A and B are fundamentally equal, it should make no difference when the output of one system is exchanged with the output of the other. To this end, the test is performed by swapping results from system A with B and \textit{vice versa}, aggregating the permuted samples, and repeating the process in an effort to estimate the sample statistic distribution. The t-test is then run on this final distribution.
For non-trivial data sizes, however, there are too many data permutations for practical computation. Thus, the sample statistic distribution is approximated by drawing random data permutations a fixed number of times.
We applied the test two-tailed at the topic level of individual ablation runs. We gathered the results of all topics for a specific run and compared these numbers with the respective outcomes of the baseline system.

\section{Explored Features}
\label{sec:features}

\subsection{BM25 Parameters}
\label{sec:bm25}

The Okapi BM25 ranking function is commonly employed in modern search engines due to its success in early TREC years. It has its roots in a probabilistic ranking approach, trying to answer the question of how likely a document $d$ is relevant given a particular query $q$. The essential factors of the BM25 scoring formula, given in Equation~(\ref{eq:bm25}), are the term frequency of terms $t$ in document $d$, denoted $tf_{td}$, the document frequency of $t$, denoted $df_t$, the total number of documents in the collection, denoted $N$, and the normalized document length $\frac{L_d}{L_{ave}}$.

\begin{equation}
\label{eq:bm25}
BM25(q,d) := \sum_{t \in q} \log \frac{N - df_t + \frac{1}{2}}{df_t + \frac{1}{2}} \times \frac{(k_1 + 1)tf_{td}}{k_1((1-b)+b\frac{L_d}{L_{ave}})+tf_{td}}
\end{equation}

To control the impact of the term frequency and the document length normalization, two hyperparameters, $k_1$ and $b$, were originally introduced, prone to adaption to a particular task and dataset. While existing search engine libraries often come with 
default values, there is no guarantee that such settings will perform well or even optimal on a given set of documents \cite{Robertson10}. Still, for the ablation analysis performed in this paper, we chose default parameters as set by \ac{ES}: $k_1:=1.2$ and $b:=0.75$.

\subsection{Query Type}
\label{sec:dismax}

\paragraph{Expansions}
We took into account the usage of \emph{dis\_max} vs. disjunctive composite queries for the disease and gene topic expansion terms.
The respective expansion terms constitute alternatives for each other, i.e., they rephrase the topic by plausible term variants. These lexical alternatives are often formulated as Boolean disjunctions. Yet, the disjunctive approach rewards many mentions of many alternative terms in a document which is not the main goal. 

As an alternative, we evaluated the potential of the \emph{dis\_max} query.
In their most simple form, disjunction max (\emph{dis\_max}) queries score a range of subqueries independently and ultimately output the score of the highest-scoring subquery. In this way, they are able to express alternatives in a more subtle way: it is not about how many subqueries match, but about the best match from a set of equally valid options (see Equation~\ref{eq:dis_max}, where $\{q_1,\dots,q_n\}$ is a set of alternative subqueries and the function $score$ returns a ranking score for the given query).

\begin{equation}
\label{eq:dis_max}
    dis\_max := max[score(q_1), score(q_2),\dots, score(q_n)]
\end{equation}

\paragraph{Multi-word}
The way in which the terms were searched across multiple fields could be set to a \emph{phrase} match (matches query terms in close proximity in the running text) or a bag-of-words (BoW) approach.

\subsection{Weighting Schema}
\label{sec:fieldweights}

The query layout illustrated in Figures~\ref{img:query} and~\ref{img:diseasequery} exhibits a number of numerical weights. Some of them directly apply to the subqueries (e.g., disease query as a whole, disease synonyms, and gene synonyms), others to the index fields (e.g., title and abstract).

\paragraph{Clauses}
The subquery weights can be used to balance the different subqueries against each other. The disease and gene aspects of the topics are of high importance while additional keywords are amplifying relevance signals that should not curtail
the main topic aspects in comparison. By default, all subquery weights are set to 1.00. Numbers greater than 1.00 cause a scoring boost to the respective subquery, while lower numbers reduce its influence on the final score. The weighted subquery score is the product of the original subquery score and the weight in \ac{ES}.

\paragraph{Fields}
A common feature of search engines is to assign higher weights to fields for which a higher \textit{a priori} probability of the occurrence of terms with relevance to the topic can be expected (e.g., titles). However, the concrete value of a weight to be assigned to a particular field is subject to experiments. The neutral value for field weights is again 1.00 and final scores are comprised of the product of the original score and the weight.

\subsection{Query Expansion}

\paragraph{Disease Expansion}
We exploited the \ac{UMLS}\footnote{\url{https://www.nlm.nih.gov/research/umls/index.html}} in version 2019AA for our disease expansion strategy. Disease expansion was implemented for synonymy, hypernymy, and preferred terms. We only used non-suppressed English concepts and terms.
For synonym expansion, we compiled the list of associated terms for each \ac{CUI} and collected all terms as synonyms that belonged to the same \ac{CUI} as the original term.
For hypernyms, we first collected the parent-relationships between \acp{CUI}. To obtain the hypernyms for a given term, we mapped the term to all \acp{CUI} listing the term as one of its synonyms as described for synonym expansion. For each such \ac{CUI}, we retrieved the direct parent \ac{CUI}. Finally, we collected all terms belonging to those parent \acp{CUI} as hypernyms for the input term.
To derive preferred disease terms, we first obtained the \acp{CUI} of the input term. Then, we retrieved the terms marked as \emph{preferred} in the \ac{UMLS} with regards to the input \acp{CUI}. As those are often several terms, we applied a majority vote to obtain the final preferred term.

Moreover, an expansion feature used by \trecpm participating teams is the addition of the term \emph{``solid''} for disease names that denote neoplasms classified as solid tumors. For this purpose, we manually compiled a list of 
disease topics denoting solid tumors and used it for expansion.

\paragraph{Gene Expansion}
We expanded genes to their synonyms and descriptions leveraging the data provided by the
NCBI.\footnote{\url{ftp://ftp.ncbi.nlm.nih.gov/gene/DATA/GENE\_INFO/Mammalia/Homo\_sapiens.gene\_info.gz}}
Additionally, we evaluated an extra match on the gene family (e.g., BRCA2 $\rightarrow$ BRCA) automatically extracted with the pattern below:
\[
\texttt{([0-9]\{1,2\}[A-Z]\{0,2\}|R[0-9]\{0,1\})\$}
\]

\subsection{Stop Word Filtering}
\label{sec:queryfiltering}

Participants of the \ac{TREC-PM} series observed a positive effect on \ac{IR} scores when domain-specific stop words were removed from the query input terms and their expansions.
Their inclusion promotes unintended hits in the result lists because they are either too general (e.g., \textit{``cancer''}) or too specific/thematically inadequate (e.g., \textit{``microsatellite''}).  Since these words come from a closed set of terms, we decided not to search for an optimal subset of stop words in this work but to use the whole list or no stop words at all.
Table~\ref{tab:query-stoplist} presents the aggregated candidate list of domain stop words extracted from participant papers.

\begin{table}[h!]
    \caption{The candidate list of domain stop words.}
    \label{tab:query-stoplist}
    \centering
    \begin{tabular}{c}
    \toprule
        adenocarcinoma amplification by ca cancer carcinoma\\
        caused cell cells defect disorder due essential familial\\
        for function instability malignant microsatellite mucosal neoplasm\\
        nerve of primary rearrangement stage the to tumor tumour with \\
        \bottomrule
    \end{tabular}
\end{table}

\subsection{Keyword Boosting}
\label{sec:keywords}

In order to boost results related to \ac{PM},\footnote{One of the most important goals in \ac{TREC-PM} is to distinguish \ac{PM}-relevant documents from non-\ac{PM}-relevant ones. If one had a classifier to reliably judge documents for this relevance decision, it would be possible to exclude a large set of candidates from the result lists. Such classifiers have been tested \cite{oleynik2018hpi,Zhou2018TeamCA}, yet did not show result score enhancements over non-classifier approaches.}
we collected positive and negative keywords from papers from well-performing participant teams. In our experiments, the words are independently toggled active or inactive in an effort to find the best overall keyword boosters. The candidate list employed in our work is depicted in Table~\ref{tab:pm-keywords}. As shown in Figure \ref{img:query}, keywords are added in optional subqueries with different weights for positive and negative words, $w_p$ and $w_n$, respectively. The impact on the BM25 document score is additive. Let the score of document $d$ without keyword boosting be $s_d$, the set of active positive keywords be $\mathcal{P}$ and the set of active negative keywords be $\mathcal{N}$. Then, the document score with applied keyword boosting in \ac{ES} 5.4 is:%
\footnote{Note that we use a negative weight for the negative keywords. This was possible in \ac{ES} 5.4, but support for negative weights was removed in newer versions.}

\begin{equation}
score(d) := s_d+\sum_{p\in \mathcal{P}} BM25(p,d)\cdot w_p + \sum_{n\in \mathcal{N}} BM25(n,d)\cdot w_n
\end{equation}

\begin{table}[h!]
    \caption{The candidate list of \ac{PM}-topic boosting keywords. Underlined words were used as starting points in the configuration search.}
    \label{tab:pm-keywords}
    \centering
    \begin{tabular}{c}
    \toprule
    \textbf{Positive} \\
    \midrule
         base \underline{clinical} cure dna efficacy \underline{gefitinib} gene \\
         genotype \underline{Gleason} heal healing malignancy \underline{outcome} \\
         patient personalized prevent prognoses \underline{prognosis} \\
         \underline{prognostic} prophylactic prophylaxis recover recovery \\
         recurrence \underline{resistance} study surgery \underline{survival} survive \\
         target \underline{targets} therapeutic therapeutical \underline{therapy} \underline{treatment}  \\
         \midrule
         \textbf{Negative} \\
         \midrule
         \underline{tumor} \underline{cell} \underline{mouse} \underline{model} \underline{tissue} \underline{development} \\
         \underline{specific} \underline{staining} \underline{pathogenesis} \underline{case} \underline{dna} \\
         \bottomrule
    \end{tabular}
\end{table}

In an attempt to simply remove false positive search results from the \textit{melanoma}-related documents (of which there are 28 topics overall, making up a large portion of the available topics) we also evaluated a exclusion subquery for the term \emph{``non-melanoma''}.
We finally evaluated promoting documents with a match on genes extracted by the gene tagger as described in Section~\ref{sec:data}.

\section{Results}

\subsection{Configuration Search}
\label{sec:results-smac}

Table~\ref{tab:optimal} shows the optimal values found using \ac{SMAC} as described in Section~\ref{sec:smac}.
For continuous parameters (e.g., BM25 hyperparameters), we report the mean value and the standard deviation across the ten cross-validation splits;
for binary features (e.g., stop word filtering), we report the number of splits in which the enabled feature was found to be optimal;
for word lists (e.g., positive keyword boosting), we report the words present in the top-1 majority of splits.
Due to space restrictions, we display the optimal value of the overall disease and gene clauses only and refer the reader to our data archive (see Section \ref{sec:conclusion})
for additional data comprising all boosting words and the complete weighting schema.

\begin{table}
  \caption{Configuration search: optimal values.}
  \label{tab:optimal}
  \begin{tabular}{l>{\raggedright\arraybackslash}p{2.5cm}>{\raggedright\arraybackslash}p{2.5cm}}
    \toprule
    Configuration & \ac{BA} & \ac{CT} \\
    \midrule
    BM25 Parameters\\
    \quad $b$ & $0.40 \pm 0.133$ & $0.72 \pm 0.313$ \\
    \quad $k_1$ & $1.11 \pm 0.126$ & $0.21 \pm 0.128$ \\
    \midrule
    Query type \\
    \quad Expansions & 10/10: dis\_max & 9/10: dis\_max \\ %
    \quad Multi-word & 10/10: phrase (disease synonym, gene synonym), 10/10: bag-of-words (gene topic, disease) & 10/10: bag-of-words (gene topic) \\
    \midrule
    Weighting schema \\
    \quad Disease clause & $1.59 \pm 0.314$ & $2.17 \pm 0.483$ \\
    \quad Gene clause & $1.58 \pm 0.658$ & $2.10 \pm 0.603$ \\
    \midrule
    Disease expansion \\
    \quad Preferred term & 8/10 & 6/10 \\    %
    \quad Synonyms & 9/10 & 8/10\\    %
    \quad Hypernyms & 4/10 & 1/10\\   %
    \quad Solid tumor & 1/10 & 10/10\\
    \midrule
    Gene expansion \\
    \quad Synonyms & 10/10 & 10/10\\ %
    \quad Description & 4/10 & 4/10\\  %
    \quad Family & 4/10 & 10/10\\   %
    \midrule
    Stop word filtering & 10/10 & 10/10 \\
    \midrule
    Keyword boosting \\
    \quad Positive & 10/10: clinical, outcome, prognosis, prognostic, survival, therapy, treatment & 8/10: prognosis, prognostic, resistance, study, targets, therapeutical \\
    \quad Negative & 10/10: dna, staining & 10/10: cell, specific \\
    \quad Non-melanoma & 6/10 & 5/10 \\
    \quad Gene tagger & $1.40 \pm 0.409$ & $1.21 \pm 0.722$ \\
    \bottomrule
\end{tabular}
\end{table}

We observe from the \ac{BA} results in Table~\ref{tab:optimal} that gene synonyms, stop word filtering, seven positive keywords and two negative keywords were selected by the optimization algorithm in all ten cross-validation splits.
The optimal mean value found for the BM25 $b$ parameter (0.40) was smaller than the default (0.75), which demonstrates a reduced importance of text length normalization for biomedical abstracts.
Finally, the mean weight computed for disease and gene clauses was similar (1.59 and 1.58, respectively) and around 60\% above the default value of 1.00.

Further data presented in the \ac{CT} column show that the solid tumor rule, gene synonyms, gene family, stop word filtering, and two negative keywords were automatically chosen in all ten splits.
The optimal mean value found for the BM25 $k_1$ parameter (0.21) was smaller than the default (1.20), which indicates a lower saturation point for term frequency.
Finally, the average weight for disease and gene clauses was similar (2.17 and 2.10, respectively) and about double as high as the default value (1.00).

\begin{figure}[t]
  \centering
  \includegraphics[width=\linewidth]{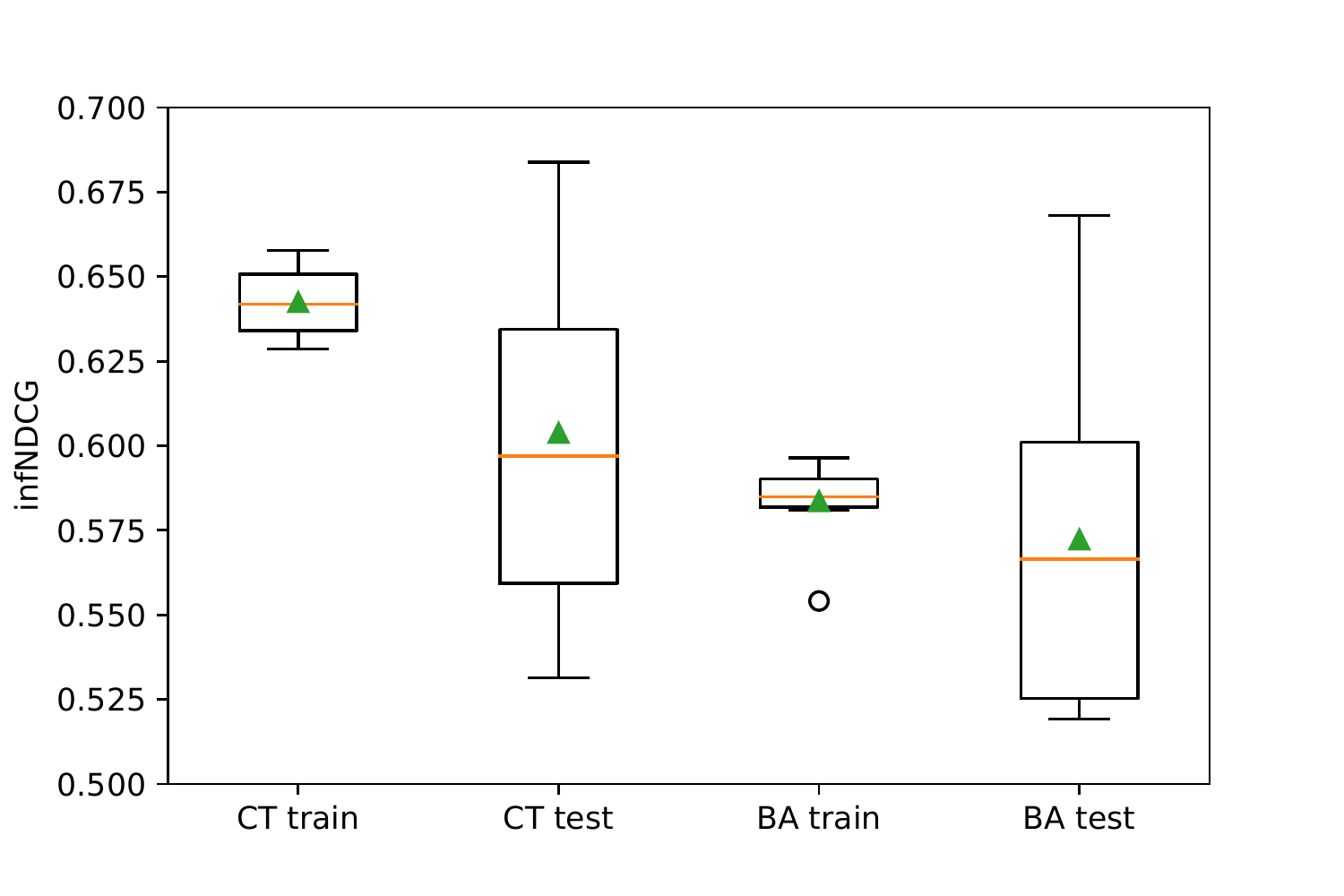}
  \caption{Final result of the parameter optimization runs on the training and test partitions.}
  \label{fig:smac_boxplot}
\Description{Figure 4. Fully described in the text.}
\end{figure}

\begin{figure}[b!]
  \centering
  \includegraphics[width=\linewidth]{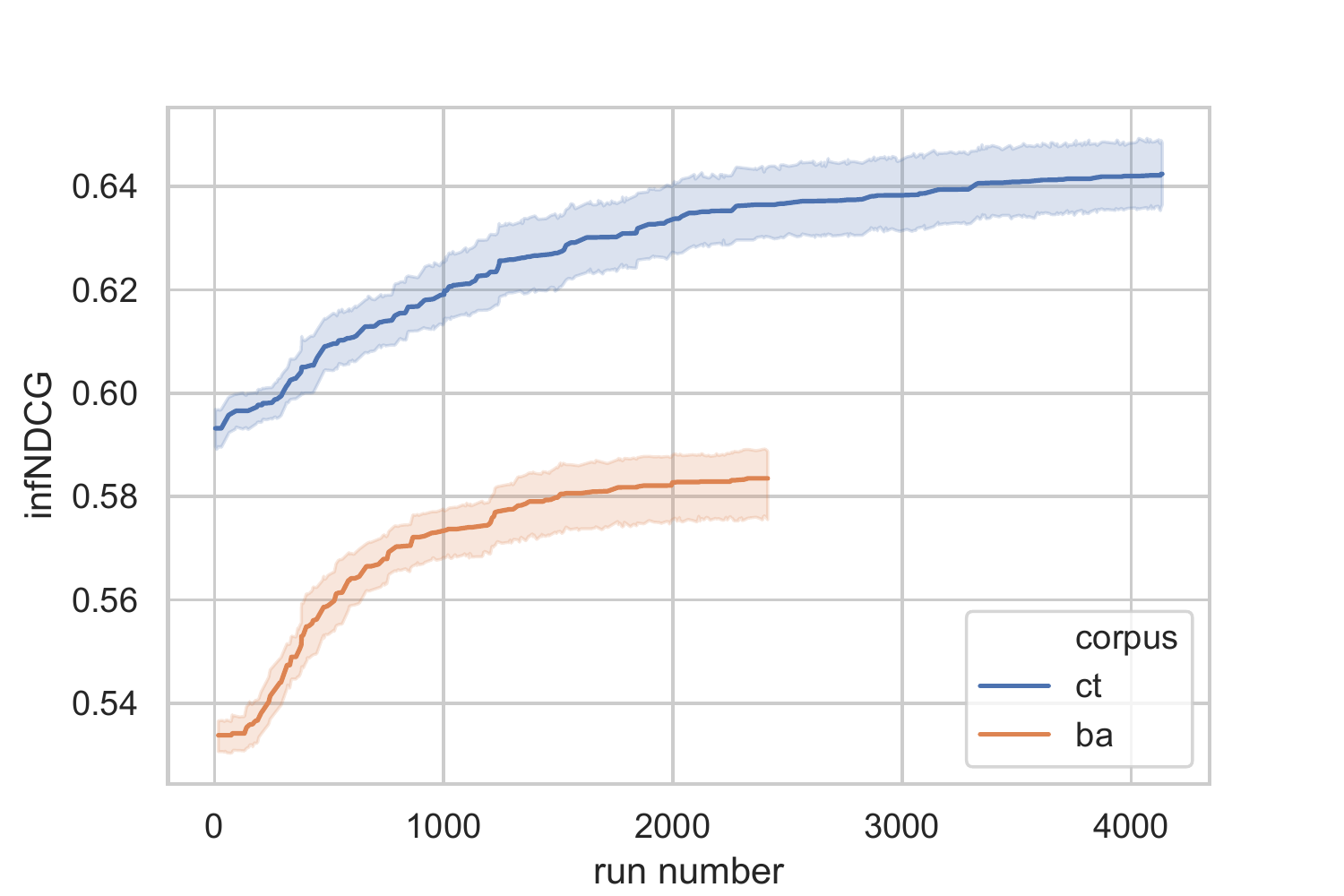}
  \caption{Progression of the \ac{SMAC} parameter optimization for \ac{BA} and \ac{CT}. The lines represent average values of the ten cross-validation splits. Additionally, the 95\% confidence interval is shown (range of lighter coloring).}
  \label{fig:smac_ctba}
  \Description{Figure 5. Fully described in the text.}
\end{figure}

Figure~\ref{fig:smac_boxplot} shows a boxplot of infNDCG metrics obtained when applying the optimal parameters for each cross-validation split in the corresponding training and test sets.
As expected, values obtained in the test splits are lower and with a larger variance 
than the ones obtained in the training splits.
On top of each boxplot, a green triangle represents the mean value; we used the test set mean values as a baseline for the ablation studies described in Section~\ref{sec:ablationresults}.
The optimal mean value obtained via \ac{SMAC} was 0.5732 and 0.6071 for the \ac{BA} and \ac{CT} tasks, respectively.

We let \ac{SMAC} run for a total of 67,776 parameter configurations, 24,755 for \ac{BA} and 43,021 for \ac{CT}.
Since the \ac{CT} data is much smaller than \ac{BA}, the evaluations ran quite a bit faster.
In these runs, SMAC found a total of 201 and 297 configurations that enhanced the previous best configuration for  \acp{BA} and \acp{CT}, respectively.

As can be seen in Figure~\ref{fig:smac_ctba}, the parameter optimization on the  \acp{BA} quickly produced better-performing configurations in the first 1,000 optimization runs. After this mark, the slope of the progress curves declines and is nearly flat from 1,500 onwards. The gain in terms of \ac{infNDCG} was between 4.00\% and 6.00\% on different cross-validation splits.
Conversely, the progression curve for \acp{CT} begins to flatten around the 2,000 run mark, with performance gains only  between 3.00\% and 6.00\% for different splits.

\subsection{Ablation Study}
\label{sec:ablationresults}

\begin{table}%
\caption{Ablation study: impact of individual system features for \ac{BA} (* denotes $p < 0.05$, ** denotes $p < 0.01$, and *** denotes $p < 0.001$).}
\begin{center}
\begin{tabular}{llc}
\toprule
Configuration & infNDCG & Difference \\
\midrule
Optimized model (baseline) & 0.5732  \\
\midrule
BM25 Parameters\\
\quad$-b$ & 0.5641* & $-1.58\%\dagger$ \\
\quad$-k_1$ & 0.5724 & $-0.15\%$ \\
\midrule
Query type \\
\quad$-$Expansions & 0.5335*** & $-6.93\%\dagger$ \\
\quad$-$Multi-word & 0.4841*** & $-15.54\%\dagger$ \\
\midrule
Weighting schema \\
\quad$-$Fields & 0.5486*** & $-4.29\%\dagger$ \\
\quad$-$Clauses & 0.5112*** & $-10.82\%\dagger$ \\
\midrule
Disease expansion \\
\quad$-$\emph{Everything} & 0.5727 & $-0.09\%$ \\
\quad$-$Preferred term & 0.5597*** & $-2.36\%\dagger$ \\
\quad$-$Synonyms & 0.5581** & $-2.64\%\dagger$ \\
\quad$-$Solid tumor & 0.5770** & $+0.65\%$ \\
\quad$+$Hypernyms & 0.5020*** & $-12.43\%$ \\
\midrule
Gene expansion \\
\quad$-$\emph{Everything} & 0.5594* & $-2.41\%$ \\
\quad$-$Synonyms & 0.5569** & $-2.85\%\dagger$ \\
\quad$-$Description & 0.5736 & $+0.06\%$ \\
\quad$-$Family & 0.5698** & $-0.59\%$ \\
\midrule
$-$Stop word filtering & 0.5033*** & $-12.21\%\dagger$ \\
\midrule
Keyword boosting \\
\quad$-$Positive & 0.5231*** & $-8.75\%\dagger$ \\
\quad$-$Negative & 0.5703 & $-0.51\%$ \\
\quad$-$Non-melanoma & 0.5735 & $+0.05\%$ \\
\quad$-$Gene tagger & 0.5566*** & $-2.90\%$ \\
\midrule
Reduced model (marked with $\dagger$) & 0.5662 & $-1.22\%$ \\% with disease and gene synonyms:
\bottomrule
\end{tabular}
\end{center}
\label{tab:bafeatureablation}
\end{table}

\begin{table}%
\caption{Ablation study: impact of individual system features for 
\ac{CT} (* denotes $p < 0.05$, ** denotes $p < 0.01$, and *** denotes $p < 0.001$).}
\begin{center}
\begin{tabular}{llc}
\toprule
Configuration & infNDCG & Difference \\
\midrule
Optimized model (baseline)  & 0.6071 \\
\midrule
BM25 Parameters\\
\quad$-b$ & 0.6093 & $+0.37\%$ \\
\quad$-k_1$ & 0.5855 & $-3.65\%$ \\
\midrule
Query type \\
\quad$-$Expansions & 0.5934 & $-2.25\%$ \\
\quad$-$Multi-word & 0.5521** & $-9.05\%\dagger$ \\
\midrule
Weighting schema \\
\quad$-$Fields & 0.6034 & $-0.60\%$ \\
\quad$-$Clauses & 0.5158*** & $-15.03\%\dagger$ \\
\midrule
Disease expansion \\
\quad$-$\emph{Everything} & 0.5344*** & $-11.61\%\dagger$ \\
\quad$-$Preferred term & 0.6067 & $-0.06\%$ \\
\quad$-$Synonyms & 0.5997 & $-1.21\%$ \\
\quad$-$Solid tumor & 0.5718* & $-5.81\%$ \\
\quad$+$Hypernyms & 0.6073 & $+0.05\%$ \\
\midrule
Gene expansion \\
\quad$-$\emph{Everything} & 0.5928 & $-2.35\%$ \\
\quad$-$Synonyms & 0.5775 & $-4.87\%\dagger$ \\
\quad$-$Description & 0.6068 & $-0.04\%$ \\
\quad$-$Family & 0.5771 & $-4.94\%$ \\
\midrule
$-$Stop word filtering & 0.5762 & $-5.09\%\dagger$ \\
\midrule
Keyword boosting \\
\quad$-$Positive & 0.6076 & $+0.09\%$ \\
\quad$-$Negative & 0.6091 & $+0.34\%$ \\
\quad$-$Non-melanoma & 0.6011 & $-0.98\%$ \\
\quad$-$Gene tagger & 0.6071 & $-0.00\%$ \\
\midrule
Reduced model (marked with $\dagger$) & 0.5962 & $-1.80\%$ \\
\bottomrule
\end{tabular}
\end{center}
\label{tab:ctfeatureablation}
\end{table}

Tables~\ref{tab:bafeatureablation} and~\ref{tab:ctfeatureablation} depict the \ac{infNDCG} metrics found by independently disabling ($-$) fine-tuned features in our ablation study for \ac{BA} and \ac{CT}, respectively.
Since disease hypernyms were not chosen by the optimization algorithm, we report metrics as if they were re-enabled ($+$) only in this case.
For continuous features (i.e., BM25 parameters and query weighting schema), disabling a configuration means setting it to the default value and effectively disabling any fine-tuning.
For query type, disabling means setting it to the default configuration (disjunction for expansions and BoW for multi-word expressions, see Table~\ref{tab:ranges}).
We also compared the value in each row to the reference optimal configuration discovered in the previous section and present the significance value of that difference.

The results show that the phrase query type plays the most important role for \ac{BA}; set to the BoW default, it is responsible for a 15.54\% drop in \ac{infNDCG} (from 0.5732 to 0.4841, $p < 0.001$).
This is closely followed by stop word filtering (12.21\% drop, $p < 0.001$), default clause weights (10.82\% drop, $p < 0.001$), removal of positive keyword boosting (8.75\% drop, $p < 0.001$), and a disjunction query type for expansions (6.93\% drop, $p < 0.001$).
Furthermore, the re-addition of disease hypernyms leads to a drop of 12.43\% in infNDCG ($p < 0.001$), an optimal configuration previously found by \ac{SMAC} (interestingly, \citet{Stokes08} come up with a similar negative result for expansion by hypernyms, yet for the general field of genomics).
Lastly, we found small, positive gains when disabling some features. Of those, only the deactivation of the solid tumor expansion is statistically significant, yet with little effect on the retrieval score. We attribute these observations to random differences between training and test data.

The results using \ac{CT} data show a similar scenario.
Here, the most important parameters are clause weights, where switching to default values leads to a drop of 15.03\% in \ac{infNDCG} (from 0.6071 to 0.5158, $p < 0.001$). They are followed by disease expansion as a whole that accounts for a drop of 11.61\% ($p < 0.001$), the phrase query type --- which, once set back to BoW, leads to a drop of 9.05\% ($p < 0.01$) ---, solid tumor rule (5.81\% drop, $p < 0.05$), gene synonyms (4.87\% drop), and stop word filtering (5.09\% drop).
Similar to \ac{BA}, we also found small,  statistically not significant, positive gains when disabling some features (e.g., positive keyword boosting).

We finally proposed our reduced model using only the features mentioned above (marked in Tables~\ref{tab:bafeatureablation} and~\ref{tab:ctfeatureablation}  with a $\dagger$).
The resulting test performance was 0.5662 \ac{infNDCG} ($-1.22\%$) for \ac{BA} and 0.5962 ($-1.80\%$) for \ac{CT}, both not significantly different from the best reference configuration ($p > 0.05$).

\section{Discussion}\label{sec:disc}

Overall, the two \trecpm collections show opposite behavior patterns, mostly probably due to their different sizes (cf.\ similar observations by \citet{Ghawi19} pointed out in Section \ref{sec:rel}).
Since the amount of \acf{BA} documents is much larger than \acf{CT}, the \ac{BA} retrieval mechanism benefits more from query boosting (especially of positive keywords) and stop word filtering to prioritize relevant documents.
Conversely, \ac{CT} takes more advantage from query expansion mechanisms (especially for \textit{disease}) that help overcome recall issues attributed to the small collection size.

Differences in document architecture may also play an important role.
While \ac{BA} documents are mostly unstructured and contain precise pieces of information, \acp{CT} are fully structured and broader in scope.
Hence, \acp{BA} benefit more from strategies that improve precision, while approaches that increase recall help \acp{CT} the most.

It is unclear whether size and document structure are confounded and thus additional work is required to disentangle these factors.
For instance, one could down-sample the \textsc{PubMed} collection to the same size of \ac{CT} in order to investigate whether the effect is mitigated or even subsides completely. 
Likewise, one could make \ac{BA} documents rarer by
removing documents from the collection and observing the impact of query expansion in this scenario.

\trecpm data show high variability among topics, not only due to their non-randomized descriptions, but also due to conflicting annotations.
This hinders supervised approaches like automated classifiers or \acf{LTR} (applied by top systems in 2018 \citep{oleynik2018hpi} and 2019 \citep{faessler2019julie}) because not enough dense data is available for training and validation.
We tried to overcome this limitation by stratifying topics per disease and gene and exploring the system as a black box model, in which biomedical particularities are not under our control.
Future work is needed to explore how the general trends observed in this study are reproducible in individual \trecpm editions or even topic-wise.

We also did not explore the impact of different ranking functions such as BM25F and BM25+, nor did we consider neural approaches based on deep learning.
For instance, BM25F may have a beneficial effect on system effectiveness owing to the structured nature of \acp{CT}.
Nonetheless, Okapi BM25 is considered a \textit{de facto} standard known to provide optimal results and thus is used by default in the underlying library \textsc{Lucene}.

Our study ran \ac{SMAC} for a finite amount of time and did not control the convergence of the found parameters that led to an optimal configuration.
We nonetheless considered alternative (and simpler) solutions in our ablation study in an effort to overcome bias.
Further investigations are required to better understand the behavior of this specific hyperparameter space and the impact of local optima in the global solution.

\section{Conclusion}
\label{sec:conclusion}

Even after many decades of information retrieval evaluation research, finding optimal choices of features and parameter settings to construct high-performance document retrieval systems has remained a challenging problem.
We found that configurations found by parameter optimization can reach an infNDCG of 0.5732 and 0.6071 on previously unseen data for the biomedical abstracts and clinical trials tasks, respectively.
Such values are compatible with values obtained by top-performing systems on every \trecpm edition.
We further believe our global optimization approach is superior to local optimization strategies that may not take unknown interdependencies among features into account.

We described the opposite behavior of the two collections with regards to query expansion and boosting.
While biomedical abstracts mostly benefit from positive keyword boosting and queries that maximize disjunction, clinical trials are aided by disease expansion (especially the solid tumor rule) and gene expansion (especially synonyms and the gene family rule).
Moreover, both datasets require optimal multi-word query types, fine-tuned clause weights, and stop word filtering (clinical trials to a lesser degree though).

With these results in mind, we proposed reduced models that can retain 98\% of the optimal retrieval scores for \pubmed and \ctdotorg, but are simpler to implement and maintain.
We believe these reduced configurations could be used in future works as an effective baseline, while research could focus either on overcoming the challenges revealed by features with a small impact on metrics or on novel directions such as neural ranking models.

The experimental code, snapshots of the search indices, and evaluation data are available at \url{https://doi.org/10.5281/zenodo.3856403} (code) and \url{https://doi.org/10.5281/zenodo.3854458} (data).

\begin{acks}
This work was supported by the BMBF within the SMITH project under grant 01ZZ1803G.
\end{acks}

\bibliographystyle{ACM-Reference-Format}
\bibliography{literature}

\end{document}